\documentclass[10pt,conference]{IEEEtran}
\usepackage[T1]{fontenc}
\usepackage{amsmath,amssymb}
\usepackage{cite}
\usepackage{graphicx}
\usepackage{booktabs}
\usepackage{stfloats}
\usepackage{placeins}
\usepackage{float}
\usepackage{url}
\usepackage{algorithm}
\usepackage{algpseudocode}
\IEEEoverridecommandlockouts

\graphicspath{{figures/}} 

\title{Synergizing Data-Driven Learning and Physical Rendering for Radio Map Prediction}
\author{
\IEEEauthorblockN{
Junyu Jiang\textsuperscript{1,2}, Jing Wang\textsuperscript{3}, Yantao Wang\textsuperscript{1}, Xinhao Li\textsuperscript{1,2}, Yi Zhang\textsuperscript{1}, Guangxu Zhu\textsuperscript{1}, Bin Wang\textsuperscript{1}\thanks{Corresponding author: wangbin@sribd.cn.}
}
\IEEEauthorblockA{
\textsuperscript{1}Shenzhen Research Institute of Big Data, Shenzhen, China\\
\textsuperscript{2}The Chinese University of Hong Kong, Shenzhen, China\\
\textsuperscript{3}China Academy of Information and Communications Technology, Beijing, China
}
}
\begin{document}
\maketitle
\begin{abstract}
Accurate urban radio map prediction is challenging due to complex building geometries and sophisticated electromagnetic interactions. To address this, we propose a data- and physics-driven framework that decouples wave propagation from environmental feature extraction. The core idea is to predict a sparse set of low-dimensional Secondary Reradiation Points (SRPs), which, together with a closed-form renderer deduced from electromagnetic theory, yields the final radio map. By learning these SRPs end-to-end rather than directly regressing the high-dimensional map, the neural network focuses on a compact physical parameterization, significantly improving accuracy and generalization. Numerical experiments demonstrate that our framework outperforms all seven baselines and consistently yields the lowest errors across all building-coverage quintiles, underscoring its robustness in dense, heavily obstructed urban environments.
\end{abstract}
\begin{IEEEkeywords}
Radio map prediction, secondary reradiation point, closed-form rendering, flow matching.
\end{IEEEkeywords}

\section{Introduction}
Radio maps characterize the spatial distribution of received signal power and serve as a fundamental tool for wireless coverage assessment, network planning, and system optimization~\cite{romero2022radiomap}. In complex urban environments, direct-path contributions can be readily approximated from the known transmitter--receiver geometry, carrier frequency, and antenna patterns~\cite{itur2024p525,3GPP2024ChannelModel}. However, the aggregate effects of building-induced reflection, diffraction, and scattering are much harder to capture. Because these higher-order interactions are highly sensitive to precise geometry, accurately accounting for them typically requires detailed 3D environment modeling and computationally expensive electromagnetic solvers~\cite{Fuschini2015RayTracing}. Consequently, achieving accurate radio map prediction at a low computational cost remains a persistent bottleneck.

To address this trade-off, existing approaches generally fall into two main categories: propagation-based methods~\cite{itur2024p525,3GPP2024ChannelModel,Fuschini2015RayTracing,hoydis2023sionna,Tadik2025OpenGERT} and data-driven methods~\cite{romero2022radiomap,Levie2021RadioUNet,Jaensch2026Coverage,Jia2026RadioMamba,Li2025RMTransformer,Viet2025STORM,Zhong2026PathFinder,Wang2025CKMTransUNet,Zhang2023RMEGAN,Wang2025RadioDiff,Luo2024RMGen,Wang2025RadioDiffK2,Liu2025WiFiDiffusion,Jia2025RadioFlow}. Propagation-based approaches rely on empirical models or deterministic ray tracing to translate environmental descriptions into received-power estimates, but they heavily rely on precise 3D environment descriptions that are difficult to obtain and remain computationally demanding. To circumvent these limitations, data-driven methods have emerged, which can be further divided into regression-based and generative AI-based approaches. Regression-based methods follow the traditional supervised learning paradigm, where neural networks with massive parameters learn a conditional mean mapping from simplified scene descriptions to radio maps given abundant training data. Early architectures like RadioUNet utilize convolutional neural networks (CNNs) to extract features and directly regress the power distribution, but they struggle with long-range reflections due to limited receptive fields~\cite{Levie2021RadioUNet}. To address this, PMNet and UNetDCN enlarge the effective receptive field using dilated and deformable convolutions~\cite{Jaensch2026Coverage}. Building upon this, subsequent works introduce state-space models and spatial transformers for global context~\cite{Jia2026RadioMamba,Li2025RMTransformer,Viet2025STORM}, and further extend the prediction to multi-transmitter and beamforming-aware scenarios~\cite{Zhong2026PathFinder,Wang2025CKMTransUNet}. However, optimizing pixel-level loss inherently drives these regression models to output an over-smoothed conditional mean, failing to capture the high-frequency variations and intricate multipath structures. To recover such fine-grained details, generative AI-based methods are introduced to learn the underlying data distribution rather than a single deterministic average. Initially, RME-GAN employs a conditional generative adversarial network (GAN) to adversarially learn the true spatial signal distribution~\cite{Zhang2023RMEGAN}. As a further evolution, diffusion-based models such as RadioDiff and RM-Gen formulate the generation as a stochastic iterative denoising process, demonstrating an enhanced capability in capturing complex spatial dependencies and high-frequency variations~\cite{Wang2025RadioDiff,Luo2024RMGen}. To optimize the generative trajectory itself, flow-matching frameworks like RadioFlow learn a direct continuous path between noise and target maps~\cite{Jia2025RadioFlow}. However, pure data-driven generation often lacks physical consistency and relies heavily on dense data. To address these practical challenges, recent efforts have begun to integrate physical laws and adapt to sparse scenarios. Specifically, WiFi-Diffusion adapts the generative process to low-sampling-rate environments~\cite{Liu2025WiFiDiffusion}, and RadioDiff-\(k^2\) incorporates the Helmholtz equation to constrain the generation of multipath singularities~\cite{Wang2025RadioDiffK2}. Despite these advances, a common limitation remains across all data-driven methods: the neural network is still forced to directly generate the final high-dimensional power distribution. This paradigm inherently entangles environmental feature extraction with the implicit approximation of wave propagation inside a single black box. Such entanglement drastically enlarges the solution space and hinders the full exploitation of model capacity, making it particularly difficult to generalize to unseen scenes.

\begin{figure}[t]
    \centering
    \includegraphics[width=\linewidth]{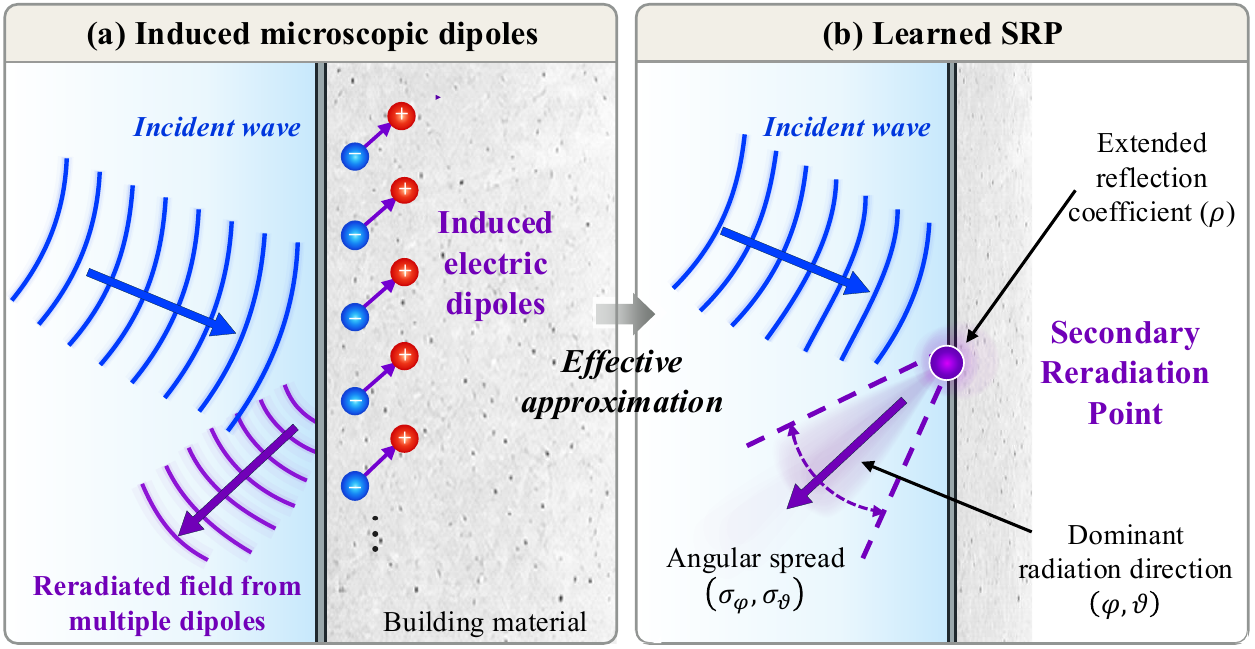}
    \vspace{-10pt}
    \caption{Illustration of the learned SRP.}
    \vspace{-10pt}
    \label{fig:srp}
\end{figure}

To resolve this entanglement, we propose a data- and physics-driven framework that explicitly synergizes data-driven learning with physical rendering. Grounded in classical electromagnetic equivalence principles~\cite{Schelkunoff1936Equivalence,Balanis2024AdvancedEM}, scattered fields can be represented by equivalent surface currents. However, rigorously evaluating these currents is prohibitive due to unknown material properties. As illustrated in Fig.~\ref{fig:srp}, we bypass this by abstracting complex scattering into a set of learned SRPs. Each SRP acts as a data-driven virtual point, parameterized by its strength, height, and directional responses, to aggregate local scattering effects. A closed-form electromagnetic renderer then computes their distance- and direction-dependent contributions, fusing them with a geometry-attenuated direct component to produce the final reference signal received power (RSRP) maps. By cascading the neural predictor with this physical renderer, the learning task is fundamentally decoupled and confined to a lower-dimensional space. Experiments demonstrate that our framework outperforms all seven baselines and consistently yields the lowest errors across building-coverage quintiles, proving its robustness in dense urban environments.

\section{System Model and Problem Formulation}
\label{sec:system_model}

\subsection{Scene and Transmitter Configurations}
\label{sec:scene_config}

We consider a set of urban scenes $\mathcal U$. Each scene $u\in\mathcal U$ is represented as an image of size \(N_x \times N_y\), where each pixel is indexed by \((i,j)\). The building geometry is represented by a height map \(H_u \in \mathbb{R}^{N_x \times N_y}\), where \(H_u[i,j]=0\) indicates no building footprint, while a positive value specifies the building height at pixel \((i,j)\). The height map serves as the condition of the flow matching model and is also used to impose constraints on the SRPs. Each transmitter configuration \(c\in\mathcal C\) is specified by
\begin{equation}
    c=(u,\boldsymbol{t}_c,P_c^{\mathrm{tx}},z_c^{\mathrm{rx}},f_c,\phi_c,\theta_c,b_c),
    \label{eq:tx_config}
\end{equation}
where $\boldsymbol{t}_c\in\mathbb R^3$ denotes the transmitter (Tx) location, $P_c^{\mathrm{tx}}$ is the Tx power in dBm, $z_c^{\mathrm{rx}}>0$ is the receiver (Rx) height, $f_c>0$ is the carrier frequency, $\phi_c$ (resp. $\theta_c$) is the azimuth (resp. tilt) angle of the Tx, and \(b_c\) is a specific beam drawn from the candidate beam set \(\mathcal B\). Specifically, for a uniform planar array with \(N_{\text{ant}}\) elements, the candidate beam set \(\mathcal B\) is generated by applying a steering vector \(\boldsymbol{w}_b \in \mathbb{C}^{N_{\text{ant}}\times 1}\) pointing towards a specific direction \((\phi_b, \theta_b)\):
\begin{equation}
\textstyle[\boldsymbol{w}_b]_n = \frac{1}{\sqrt{N_{\text{ant}}}} \exp\left(j \frac{2\pi}{\lambda} \boldsymbol{r}_n^T \boldsymbol{u}(\phi_b, \theta_b)\right),
\end{equation}
where \(\lambda\) is the carrier wavelength, \(\boldsymbol{r}_n \in \mathbb{R}^3\) is the 3D position vector of the \(n\)-th element (with \(z=0\) for a planar array), and \(\boldsymbol{u}(\phi_b, \theta_b)\) is the 3D unit directional vector. Given a scene \(u\) and a Tx configuration \(c\), the RSRP at each point of \(u\) is determined. Let \(\boldsymbol{Y}_c \in \mathbb{R}^{N_x \times N_y}\) denote the target RSRP map corresponding to \(c\). Notably, the complete multi-beam RSRP map can be obtained by simply concatenating each single-beam RSRP map along the beam dimension.

\subsection{Problem Formulation}
\label{sec:problem_formulation}

Given the building height map $H_u$ and the Tx configuration $c$, our objective is to learn a prediction model $\mathcal F_{\boldsymbol\Phi}(H_u,c)\in\mathbb R^{N_x\times N_y}$ to estimate the radio map of a scene $u$, where $\boldsymbol\Phi$ denotes the trainable parameters of $\mathcal F_{\boldsymbol\Phi}$. In particular, this prediction model is expected to generalize well to scenes and Tx configurations that are not included in the training set.

\section{Proposed Method}
\label{sec:proposed_method}

\subsection{Overview}
\label{sec:method_overview}
\begin{figure*}[!t]
    \centering
    \includegraphics[width=\textwidth]{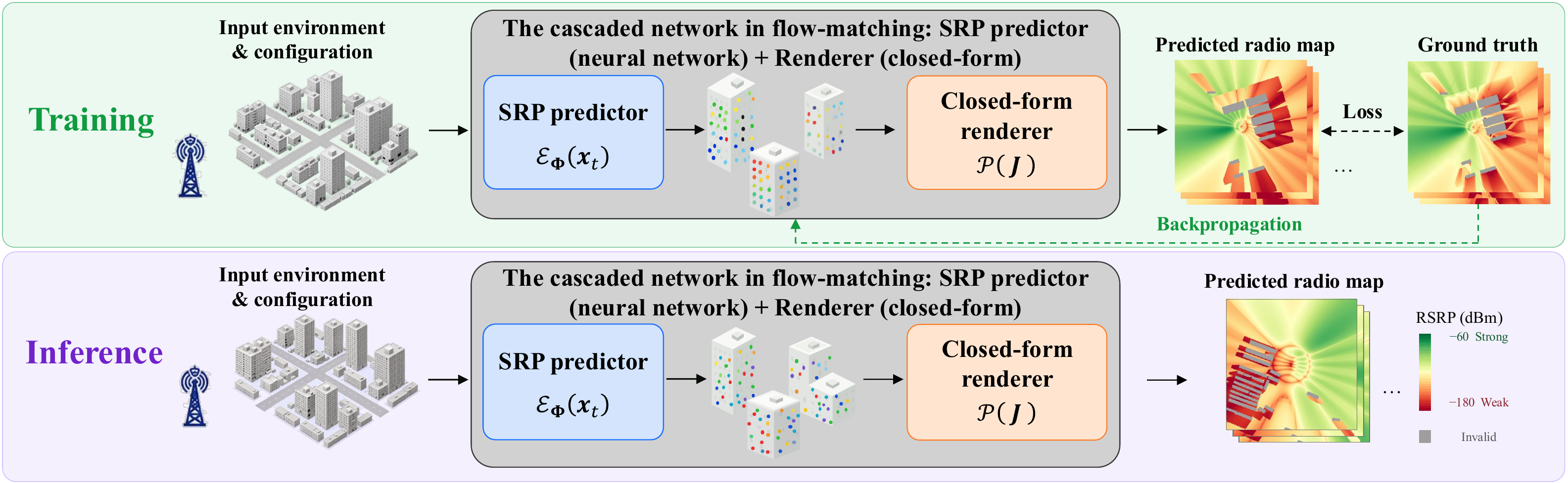}
    \vspace{-10pt}
    \caption{SRP-based radio map prediction.}
    \label{fig:esf_framework}
    \vspace{-10pt}
\end{figure*}

Our basic idea is to employ a conditional flow-matching model to generate the radio map. Instead of treating the radio map prediction problem as a traditional image generation task, we aim to learn a set of SRPs, which are then used by a closed-form renderer. By rendering only a selected subset of top-\(K\) SRPs, the network focuses on a compact physical parameterization rather than a dense image grid. The overall framework is illustrated in Fig.~\ref{fig:esf_framework}. In the rest of this section, we elaborate the building blocks of our method, i.e., the SRPs, the closed-form propagation renderer, and the flow-matching-based generative model.

\subsection{Physics-Constrained SRP Representation}
\label{sec:esf_representation}

Unlike natural RGB images governed by visual appearance, radio maps are governed by electromagnetic propagation laws. Forcing a black-box network to directly fit these maps results in an excessively large solution space. Moreover, the network cannot make full use of its capacity, because the network must simultaneously extract environmental features and implicitly approximate wave propagation. Standard networks can easily capture line-of-sight (LoS) attenuation, but struggle to fit complex multipath interactions such as reflection and diffraction. To overcome this challenge, we decouple feature extraction from physical propagation modeling: the neural network only needs to predict SRP distributions, while a closed-form propagation renderer explicitly computes wave propagation. In this perspective, the output of the neural network is confined to a lower-dimensional space, which can potentially improve model accuracy and generalization. This decoupling is physically grounded in electromagnetic equivalence principles, specifically the induction theorem and surface equivalence formulations detailed in~\cite[Sec. 5]{Schelkunoff1936Equivalence} and~\cite[Ch. 7.9--7.12]{Balanis2024AdvancedEM}. Incident waves interacting with buildings induce polarization and conduction currents that reradiate secondary fields. Under a Green-function formulation, these scattered fields can be represented by equivalent point distributions separated from the propagation operator that maps them to receiver locations. Since it is usually difficult to obtain the precise electromagnetic properties of the building blocks in real scenarios, we aim to generate a number of parameterized SRPs, including their spatial positions, radiation strength, dominant radiation directions as well as angular spreads.

\subsubsection{SRP Parameterization}
\label{sec:srp_parameterization}
For a Tx configuration $c$ in scene $u$, the SRPs are encoded into an eight-channel image \(\boldsymbol{J}_c\in\mathbb R^{8\times N_x\times N_y}\), in which
\begin{equation}
\begin{aligned}
\textstyle \boldsymbol{J}_c[i,j] = [& s_c[i,j], \rho_c[i,j], \cos\varphi_c[i,j], \sin\varphi_c[i,j], \\
& \sigma_{\varphi,c}[i,j], h_{z,c}[i,j], \vartheta_c[i,j], \sigma_{\vartheta,c}[i,j] ]^T,
\end{aligned}
\label{eq:srp_parameterization}
\end{equation}
where \(s_c[i,j]\in[0,1]\) is the probability of whether the position \((i,j)\) has an SRP, \(\rho_c[i,j]\in(0,\rho_{\max})\) is the extended reflection coefficient bounded by \(\rho_{\max}>0\), \(\varphi_c[i,j]\in(-\pi,\pi]\) is the dominant azimuth angle, \(\sigma_{\varphi,c}[i,j]>0\) is the azimuthal beamwidth, \(h_{z,c}[i,j]\in(0,1)\) is the normalized point height relative to the nearest building, \(\vartheta_c[i,j]\in[-\pi/2,\pi/2]\) is the dominant tilt angle, and \(\sigma_{\vartheta,c}[i,j]>0\) is the tilt beamwidth. The azimuth angle \(\varphi_c[i,j]\) is transformed to the polar coordinate system by constructing the unit vector \((\cos\varphi_c[i,j],\sin\varphi_c[i,j])\), which can avoid angular wraparound discontinuities. The secondary reradiation weight is defined as \(\omega_c[i,j]=s_c[i,j]\rho_c[i,j]\), which is used for selecting a small portion of SRPs. Notably, activating only a small portion of SRPs enforces spatial sparsity to prevent overfitting, which aligns with the physical fact that multipath energy concentrates on a few strong scatterers.

\subsubsection{SRP Position Constraints}
To enhance physical interpretability, the SRPs are subject to spatial and visibility constraints. First, each SRP should be located in an open area near a building, and its height must not exceed the roofline of the nearest building. Given a region radius $r_{\mathrm{reg}}>0$, the admissible set of candidate SRPs is denoted as
\begin{equation}
\textstyle \mathcal{W}_u = \{ (i,j) : H_u[i,j] = 0, \, D_{u}^{\mathrm{wall}}[i,j] \le r_{\mathrm{reg}} \},
\label{eq:esf_candidate_support}
\end{equation}
where \(D_{u}^{\mathrm{wall}}[i,j]\) is the distance from pixel \((i,j)\) to the nearest building, and the condition \(H_u[i,j]=0\) ensures that the candidate SRPs are strictly located in open areas rather than inside building footprints. The probability \(s_c[i,j]\) is forced to zero for all points outside \(\mathcal{W}_u\). Each candidate point $(i,j)\in\mathcal W_u$ is placed at height \(z_{c}[i,j]=h_{z,c}[i,j]\cdot H_{u}^{\mathrm{nst}}[i,j]\), where \(H_{u}^{\mathrm{nst}}[i,j]\) denotes the height of the nearest building to pixel \((i,j)\), and \(h_{z,c}[i,j]\in(0,1)\) is the scaling ratio predicted by the SRP predictor, ensuring that the point height remains at or below the roofline of the nearest building. 

Second, each SRP should lie in the LoS region of the Tx. To impose this constraint, we compute a binary visibility map \(V\in\mathbb{R}^{N_x\times N_y}\). The entry \(V[i,j]\) is set to 1 if there exists an LoS path between the Tx and pixel \((i,j)\), and 0 otherwise. After obtaining the raw SRP prediction \(\boldsymbol{J}_{c,t}\) (see Section~\ref{sec:flow_esf_prediction}), \(V\) is applied element-wise to mask the activation probability channel \(s_c\) within \(\boldsymbol{J}_{c,t}\) to exclude physically non-plausible SRPs.

\subsection{Closed-Form Propagation Renderer}
\label{sec:propagation_renderer}

The complete radio map is generated by a closed-form physical renderer \(\mathcal{P}\), which synthesizes both the LoS and secondary scattering contributions. The predicted RSRP is obtained according to
\begin{equation}
\begin{aligned}
\textstyle    &\widehat y_{c}[i',j'] = \big[\mathcal{P}(\boldsymbol{J}_c; H_u, c)\big]_{i',j'} \\
    &= P_c^{\mathrm{tx}} \!+ \!10\log_{10}\!\Big( \!\max\{D_{c}[i',j'; H_u] \!+ \!R_{c}[i',j'; \!\boldsymbol{J}_c,\! H_u], \epsilon\} \!\Big),
\end{aligned}
\label{eq:esf_rsrp_synthesis}
\end{equation}
where $\widehat y_{c}[i',j']$ is the predicted RSRP (in dBm) at Rx pixel \((i',j')\), $P_c^{\mathrm{tx}}$ is the transmit power, \(D_{c}[i',j'; H_u]\) is the LoS path component determined by the environment geometry \(H_u\), \(R_{c}[i',j'; \boldsymbol{J}_c, H_u]\) is the aggregated secondary reradiation contribution computed based on the predicted SRP image \(\boldsymbol{J}_c\) and geometry, and $\epsilon>0$ is a numerical floor for the logarithm. The LoS path gain is obtained via $D_{c}[i',j'; H_u]=\eta_{c}[i',j'; H_u]\,\cdot 10^{L_{c}[i',j']/10}$, where $\eta_{c}[i',j'; H_u]\in[0,1]$ is the attenuation factor determined from \(H_u\) and the Tx--Rx geometry, and $L_{c}[i',j']$ is the beam-specific free-space illumination (in dB) evaluated at the Rx point $(i', j')$, which is computed via
\begin{equation}
\textstyle L_{c}[i',j'] = G_{c}[i',j'] - \mathrm{FSPL}_{c}[i',j'],
\end{equation}
where \(G_{c}[i',j']\) is the Tx antenna gain and \(\mathrm{FSPL}_{c}[i',j']\) is the free-space path loss computed based on the Tx-Rx distance and carrier frequency~\cite{itur2024p525}. Within \(\mathcal{P}\), the secondary scattering component is computed via
\begin{equation}
\textstyle    R_{c}[i',j'] = \sum_{(i,j)\in\mathcal A_c} \omega_{c}[i,j]\cdot \kappa_{c,(i,j),(i',j')},
\label{eq:esf_source_aggregation}
\end{equation}
where \(\mathcal A_c\) is the set of all valid SRPs, \(\omega_{c}[i,j] = s_c[i,j]\rho_c[i,j]\) is the SRP weight, and \(\kappa_{c,(i,j),(i',j')}\) is the directional propagation from SRP position \((i,j)\) to receiving point \((i',j')\). According to Section~\ref{sec:srp_parameterization}, the active set \(\mathcal A_c\) is formed by selecting the \(K\) candidate SRPs with the largest positive weights \(\omega_c[i,j]\). Notably, all predicted variables in \(\boldsymbol{J}_c\) are explicitly incorporated into the computation of \(\omega_{c}\) and \(\kappa_{c}\), forming a complete physical mapping from the SRP image to the rendered power. According to the free-space attenuation model~\cite{itur2024p525} and the 3GPP directional antenna model~\cite{3GPP2024ChannelModel}, the propagation kernel \(\kappa\) has the following form:
\begin{equation}
\begin{aligned}
\textstyle    \kappa_{c,(i,j),(i',j')} &= \Big(\frac{\Delta \ell}{\ell_0}\Big)^{-\nu}
    \!\exp\!\!\Big[
    -\frac{(\Delta\varphi_{c})^2}
    {2\sigma_{\varphi,c}^2[i,j]}
    \!\Big] \\
    &\quad \times \!\exp\!\!\Big[
    -\frac{(\Delta\vartheta_{c})^2}
    {2\widetilde\sigma_{\vartheta,c}^2[i,j]}
    \!\Big],
\end{aligned}
\label{eq:esf_propagation_kernel}
\end{equation}
where \(\Delta \ell=\sqrt{d^{2}+\delta^{2}}\) is the smoothed 3D distance between the SRP and the Rx (with \(d\) being the physical distance and \(\delta>0\) a small constant preventing singularity). \(\ell_0=1\)~m is the reference distance, \(\nu\) is a distance-attenuation exponent, \(\Delta\varphi_{c}\) (resp. \(\Delta\vartheta_{c}\)) is the azimuth (resp. tilt) angular offset from the point's dominant radiation direction, and \(\widetilde\sigma_{\vartheta,c}[i,j]=\operatorname{clip}_{[\sigma_{\vartheta,\min},\sigma_{\vartheta,\max}]}(\sigma_{\vartheta,c}[i,j])\) is the clipped tilt beamwidth. The angular offsets are defined as \(\Delta\varphi_{c} = \operatorname{wrap}_{\pi}(\alpha_{c} - \varphi_{c}[i,j])\) and \(\Delta\vartheta_{c} = \beta_{c} - \vartheta_{c}[i,j]\), where \(\operatorname{wrap}_{\pi}(v)=\operatorname{atan2}(\sin v,\cos v)\) ensures angular periodicity, \(\alpha_{c}\) and \(\beta_{c}\) are the geometric departure angles from the SRP to the Rx, and \(\varphi_{c}[i,j]\) and \(\vartheta_{c}[i,j]\) are the predicted dominant radiation directions.

\subsection{Flow-Matching-Based Radio Map Generation}
\label{sec:flow_esf_prediction}

Conditional Flow Matching (CFM) provides a general framework for transforming a Gaussian distribution \(p_0(\boldsymbol{x}_0) = \mathcal{N}(\mathbf{0}, \mathbf{I})\) into a target conditional distribution \(q_{\text{data}}(\boldsymbol{x}_1|\boldsymbol{p})\). To describe this continuous and invertible transformation, an auxiliary continuous time variable \(t \in [0, 1]\) is introduced, and a family of conditional probability densities \(\{p_t(\boldsymbol{x}_t|\boldsymbol{p})\}\) is constructed such that:
\begin{equation}
\textstyle    p_0(\boldsymbol{x}_0|\boldsymbol{p}) = p_0(\boldsymbol{x}_0), \quad p_1(\boldsymbol{x}_1|\boldsymbol{p}) = q_{\text{data}}(\boldsymbol{x}_1|\boldsymbol{p}).
\end{equation}
\(\{p_t(\boldsymbol{x}_t|\boldsymbol{p})\}\) defines a conditional probability path that continuously transports probability mass from the base distribution to the target conditional distribution, which is assumed to be generated by a deterministic flow governed by the ordinary differential equation (ODE):
\begin{equation}
    {d\boldsymbol{x}_t}/{dt} = \boldsymbol{v}_t(\boldsymbol{x}_t|\boldsymbol{p}), \quad \boldsymbol{x}_0 \sim p_0(\boldsymbol{x}_0),
\end{equation}
where \(\boldsymbol{v}_t\) is a time-dependent conditional velocity field. A fundamental result from the theory of continuous normalizing flows states that the density \(p_t(\boldsymbol{x}_t|\boldsymbol{p})\) induced by the flow must satisfy the conditional continuity equation:
\begin{equation}
    \frac{\partial p_t(\boldsymbol{x}_t|\boldsymbol{p})}{\partial t} + \nabla_{\boldsymbol{x}_t} \cdot \left(p_t(\boldsymbol{x}_t|\boldsymbol{p})\boldsymbol{v}_t(\boldsymbol{x}_t|\boldsymbol{p})\right) = 0.
\end{equation}
This equation enforces the conservation of probability mass along the flow but does not uniquely determine the velocity field \(\boldsymbol{v}_t\). Infinitely many velocity fields can induce the same target probability path. A fundamental principle of CFM is to avoid implicit inference via likelihood maximization or score estimation. Instead, one explicitly specifies a probability path together with an associated target process known in closed form. While the traditional formulation trains a neural network to directly regress the velocity field \(\boldsymbol{v}_t\), an equivalent and often more effective reparameterization is to predict the target data \(\boldsymbol{x}_1\) directly. Because the target velocity is analytically linked to the target data \(\boldsymbol{x}_1\), the learning problem reduces to approximating the conditional expectation of \(\boldsymbol{x}_1\) using a neural network \(\boldsymbol{x}_{\boldsymbol{\Phi}}(\boldsymbol{x}_t, t; \boldsymbol{p})\), where \(\boldsymbol{\Phi}\) denotes the model parameters. CFM then trains \(\boldsymbol{\Phi}\) by minimizing the mean squared error loss:
\begin{equation}
\textstyle    \mathcal{L}_{\text{CFM}} = \mathbb{E}_{\boldsymbol{p}, t, \boldsymbol{x}_t} \big[ \left\| \boldsymbol{x}_{\boldsymbol{\Phi}}(\boldsymbol{x}_t, t; \boldsymbol{p}) - \boldsymbol{x}_1 \right\|_2^2 \big].
\end{equation}
When this loss is minimized, the network learns to predict the expected target data given the noisy state and condition.

We employ the flow-matching framework to realize radio map generation. In our formulation, let \(\boldsymbol{x}_1 = \mathcal{S}(\boldsymbol{Y}_c)\) denote the target normalized RSRP image, where \(\mathcal{S}(\cdot)\) is a normalization function, \(\boldsymbol{x}_t\) be the intermediate noisy state at time \(t\), and \(\boldsymbol{p} = (H_u, c)\) represent the scene conditions. The prediction process is achieved via a cascaded network, which is the concatenation of the SRP predictor and the renderer, that is,
\begin{equation}
    \widehat{\boldsymbol{Y}}_{c,t} = \mathcal{P}\big(\mathcal{E}_{\boldsymbol\Phi}(\boldsymbol{x}_t, t; \boldsymbol{p}); H_u, c\big),
\end{equation}
where the SRP predictor is realized by a Diffusion Transformer (DiT) neural network \(\mathcal{E}_{\boldsymbol\Phi}(\boldsymbol{x}_t, t; \boldsymbol{p}) \in \mathbb{R}^{8 \times N_x \times N_y}\) (corresponding to \(\boldsymbol{J}\) defined in \eqref{eq:srp_parameterization}), the renderer is denoted as \(\mathcal{P}\) (see \eqref{eq:esf_rsrp_synthesis}), and \(\boldsymbol{x}_t\) is the intermediate noisy normalized RSRP image at time \(t\).

\subsubsection{Training Procedure}
During training, we sample a time \(t \sim \mathcal{U}(0,1)\) and a noise \(\boldsymbol{x}_0 \sim \mathcal{N}(\mathbf{0}, \mathbf{I})\) to construct the intermediate state \(\boldsymbol{x}_t = (1-t)\boldsymbol{x}_0 + t\boldsymbol{x}_1\). With such a design, we only need RSRP ground truth instead of SRP ground truth for CFM training. This is essential because SRP ground truth is difficult to obtain. As opposed to standard CFM, our training loss is composed of a mean absolute error (MAE) loss and several physical constraint losses: 
\begin{equation}
\begin{aligned}
\textstyle \mathcal{L}_{\text{SRP}} = \mathcal{L}_{\text{map}} &+ \lambda_{\text{amp}}\mathcal{L}_{\text{amp}} + \lambda_{\text{sp}}\mathcal{L}_{\text{sp}} + \lambda_{\text{wall}}\mathcal{L}_{\text{wall}} \\
&+ \lambda_{\text{az}}\mathcal{L}_{\text{az}} + \lambda_{\text{el}}\mathcal{L}_{\text{el}} + \lambda_{\text{ne}}\mathcal{L}_{\text{ne}},
\end{aligned}
\label{eq:esf_training_objective}
\end{equation}
where the MAE loss is designed as
\begin{equation}
\textstyle    \mathcal{L}_{\text{map}} = \frac{\|\boldsymbol{M}_c \odot (\widehat{\boldsymbol{Y}}_{c,t} - \boldsymbol{Y}_c)\|_1}{\|\boldsymbol{M}_c\|_1},
    \label{eq:esf_map_loss}
\end{equation}
where \(\boldsymbol{M}_c \in \{0,1\}^{N_x \times N_y}\) is a binary mask whose purpose is to preclude those empty-value positions in the ground-truth sample \(\boldsymbol{Y}_c\) (which commonly occurs in ray-tracing-based simulations). The remaining six terms impose validity constraints to guide the network toward physically plausible and sparse SRP distributions. 

\begin{algorithm}[!t]
\caption{Training of the SRP Predictor}
\label{alg:esf_training}
\begin{algorithmic}[1]
\Require Training data \(\mathcal{T}\), initial parameters \(\boldsymbol\Phi, \nu\)
\Ensure Optimized parameters \(\boldsymbol\Phi^*, \nu^*\)
\Repeat
    \State Sample a batch of configurations \(c\) and ground-truth RSRP maps \(\boldsymbol{Y}_c\)
    \State Normalize the target map: \(\boldsymbol{x}_1 = \mathcal{S}(\boldsymbol{Y}_c)\)
    \State Define scene conditions: \(\boldsymbol{p} = (H_u, c)\)
    \State Sample \(\boldsymbol{x}_0 \sim \mathcal{N}(\mathbf{0}, \mathbf{I})\) and \(t \sim \mathcal{U}(0,1)\)
    \State Construct intermediate state: \(\boldsymbol{x}_t = (1-t)\boldsymbol{x}_0 + t\boldsymbol{x}_1\)
    \State Obtain predicted SRPs via \(\boldsymbol{J}_{c,t} = \mathcal{E}_{\boldsymbol\Phi}(\boldsymbol{x}_t, t; \boldsymbol{p})\)
    \State Obtain predicted RSRP map via \(\widehat{\boldsymbol{Y}}_{c,t} \!\!= \!\mathcal{P}(\boldsymbol{J}_{c,t}; H_u, c)\)
    \State Compute loss \(\mathcal{L}_{\text{SRP}}\) via \eqref{eq:esf_training_objective}
    \State Update \(\boldsymbol\Phi, \nu\) via gradient descent on \(\mathcal{L}_{\text{SRP}}\)
\Until{converged}
\State \Return \(\boldsymbol\Phi^*, \nu^*\)
\end{algorithmic}
\end{algorithm}

The amplitude limitation loss \(\mathcal{L}_{\text{amp}}\) restricts the total energy of the predicted SRPs to prevent unrealistically large reflection coefficients:
\begin{equation}
    \mathcal{L}_{\text{amp}} = \frac{1}{N_x N_y} \sum_{i,j} s_c[i,j] \rho_c[i,j].
\end{equation}
To encourage a concentrated set of strong scatterers rather than a dense, weak distribution, the sparsity loss \(\mathcal{L}_{\text{sp}}\) applies a Hoyer sparsity penalty on the activation probability map \(\boldsymbol{s}_c\):
\begin{equation}
    \mathcal{L}_{\text{sp}} = \left( \frac{\|\boldsymbol{s}_c\|_1}{\|\boldsymbol{s}_c\|_2 + \epsilon} \right)^2,
\end{equation}
where \(\|\cdot\|_1\) and \(\|\cdot\|_2\) denote the \(\ell_1\) and \(\ell_2\) norms of the flattened tensor, respectively, and \(\epsilon > 0\) is a small constant for numerical stability.
To further strengthen the physical constraint that SRPs should adhere to building surfaces, the wall proximity loss \(\mathcal{L}_{\text{wall}}\) directly penalizes active SRPs based on their distance to the nearest building:
\begin{equation}
    \mathcal{L}_{\text{wall}} = \frac{1}{N_x N_y} \sum_{i,j} s_c[i,j] \cdot D_{u}^{\mathrm{wall}}[i,j].
\end{equation}
The azimuthal constraint loss \(\mathcal{L}_{\text{az}}\) prevents excessively broad azimuthal beamwidths, which helps in maintaining sharp directional reflections:
\begin{equation}
    \mathcal{L}_{\text{az}} = \frac{1}{N_x N_y} \sum_{i,j} s_c[i,j] \cdot \max\big(0, \sigma_{\varphi,c}[i,j] - \sigma_{\varphi,\max}\big),
\end{equation}
where \(\sigma_{\varphi,\max}\) is a predefined upper bound for the azimuthal beamwidth.
The elevation constraint loss \(\mathcal{L}_{\text{el}}\) penalizes the deviation of the predicted dominant tilt angle \(\vartheta_c[i,j]\) from the geometric departure angle \(\beta_c\) to the receiver region:
\begin{equation}
    \mathcal{L}_{\text{el}} = \frac{1}{N_x N_y} \sum_{i,j} s_c[i,j] \cdot \big( \vartheta_c[i,j] - \beta_c \big)^2.
\end{equation}
Finally, to prevent the degeneracy where all activation probabilities collapse to zero (which would yield a trivial solution), the non-empty loss \(\mathcal{L}_{\text{ne}}\) serves as a soft constraint to ensure that the expected number of active SRPs is at least \(K\):
\begin{equation}
    \mathcal{L}_{\text{ne}} = \max\left(0, K - \sum_{i,j} s_c[i,j]\right)^2.
\end{equation}
The complete training procedure is summarized in Algorithm~\ref{alg:esf_training}.

\begin{algorithm}[!t]
\caption{Iterative ODE Sampling for SRP Generation}
\label{alg:esf_inference}
\begin{algorithmic}[1]
\State Initialize the total number of sampling steps \(N = 10\);
\State Load the trained parameters \(\boldsymbol\Phi^*\) and scene conditions \(\boldsymbol{p} = (H_u, c)\);
\State Sample a Gaussian noise vector \(\boldsymbol{x}_0 \sim \mathcal{N}(\mathbf{0}, \mathbf{I})\);
\State Set time grid \(\{t_i\}_{i=0}^N\) with \(t_0 = 0\), \(t_N = 1\), and step size \(\Delta t = 1/N\);
\For{\(i = 0\) \textbf{to} \(N - 1\)}
    \State Predict raw SRPs: \(\boldsymbol{J}_{c, t_i} = \mathcal{E}_{\boldsymbol\Phi^*}(\boldsymbol{x}_{t_i}, t_i; \boldsymbol{p})\);
    \State Apply wall/visibility constraints and top-\(K\) to form \(\mathcal{A}_c\);
    \State Render intermediate map: \(\widehat{\boldsymbol{Y}}_{c, t_i} = \mathcal{P}(\mathcal{A}_c; H_u, c)\);
    \State Compute target estimate: \(\widehat{\boldsymbol{x}}_1 = \mathcal{S}(\widehat{\boldsymbol{Y}}_{c, t_i})\);
    \State Update state according to \(\boldsymbol{x}_{t_{i+1}} \gets \boldsymbol{x}_{t_i} + \frac{\Delta t}{1 - t_i + \epsilon} (\widehat{\boldsymbol{x}}_1 - \boldsymbol{x}_{t_i})\);
\EndFor
\State Terminal forward pass: Obtain final SRPs \(\boldsymbol{J}_{c, 1} = \mathcal{E}_{\boldsymbol\Phi^*}(\boldsymbol{x}_{t_N}, 1; \boldsymbol{p})\) and form \(\mathcal{A}_c\);
\State Final render: \(\widehat{\boldsymbol{Y}}_{c,1} = \mathcal{P}(\mathcal{A}_c; H_u, c)\);
\State \Return The generated RSRP map \(\widehat{\boldsymbol{Y}}_{c,1}\).
\end{algorithmic}
\end{algorithm}

\subsubsection{Inference Procedure}
Upon completing the training stage, the network can predict the target data at any given time \(t\). The inference process generates the radio map by numerically integrating the ODE from \(t=0\) to \(t=1\). Specifically, we discretize the continuous time period \([0, 1]\) into \(N=10\) uniform time intervals with an interval size of \(\Delta t = 1/N\), yielding a time sequence defined as \(t_i = i \cdot \Delta t\), where \(i = 0, 1, \dots, N\). By employing the first-order Euler method and substituting the equivalent relationship between the predicted target data and the velocity field, the intermediate state \(\boldsymbol{x}_t\) is iteratively updated as follows
\begin{equation}
    \boldsymbol{x}_{t_{i+1}} = \boldsymbol{x}_{t_i} + \frac{\hat{\boldsymbol{x}}_1(t_i) - \boldsymbol{x}_{t_i}}{1 - t_i} \cdot \Delta t, \quad i = 0, 1, \dots, N-1,
    \label{eq:euler_update}
\end{equation}
where \(\hat{\boldsymbol{x}}_1(t_i) = \mathcal{S}\big(\mathcal{P}(\mathcal{A}_c(\mathcal{E}_{\boldsymbol\Phi}(\boldsymbol{x}_{t_i}, t_i; \boldsymbol{p})); H_u, c)\big)\) is the normalized target map estimated by the cascaded network, and \(\mathcal{A}_c\) is the sparse active set formed by applying wall and visibility constraints alongside top-\(K\) selection. After \(N\) iterations of the update process as shown in \eqref{eq:euler_update}, an additional terminal forward pass is executed at the final state \(\boldsymbol{x}_{t_N}\) to generate the ultimate RSRP map \(\widehat{\boldsymbol{Y}}_{c,1}\). The overall inference process is summarized in Algorithm~\ref{alg:esf_inference}.

\section{Experiments}
\label{sec:experiments}

\begin{figure*}[!t]
    \centering
    \includegraphics[width=\textwidth]{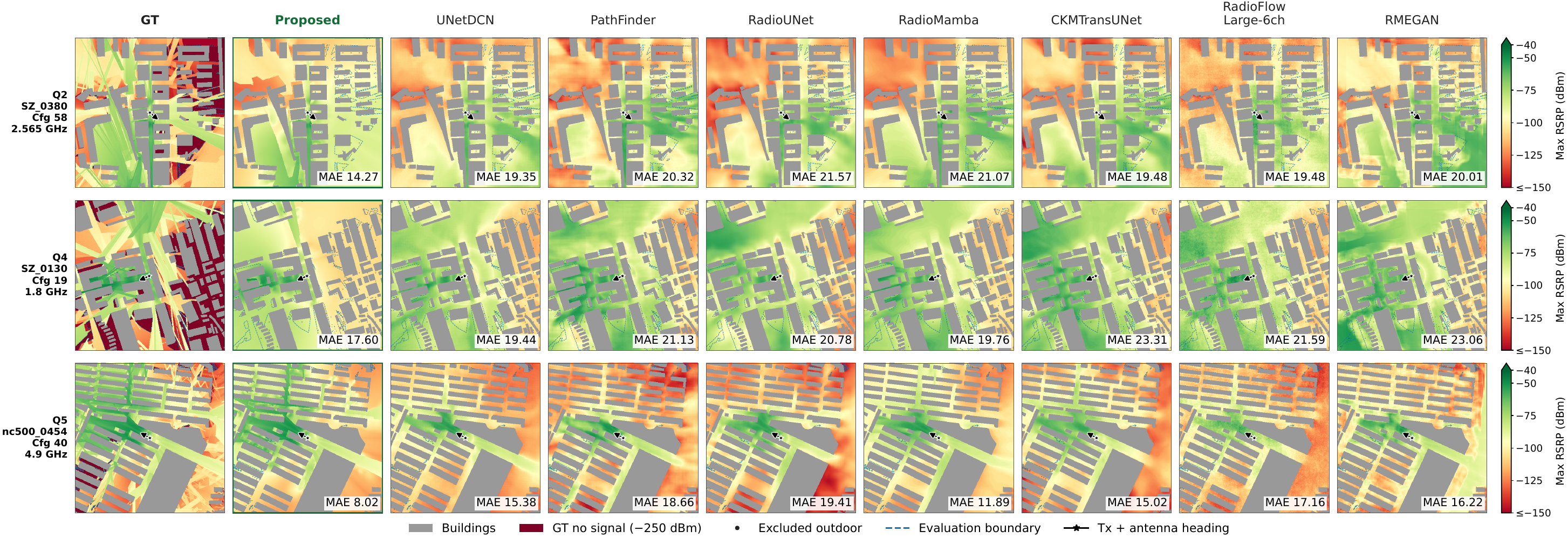}
    \vspace{-10pt}
    \caption{Maximum-RSRP maps across eight beams for three selected cases with different building coverage.}
    \label{fig:maximum_rsrp_comparison}
    \vspace{-10pt}
\end{figure*}

\subsection{Experimental Setup}
\label{sec:experimental_setup}

We evaluate on a simulated dataset of 1,000 urban scenes from Shenzhen and Nanchang (500 each), with each scene represented as a \(512\times512\) image at a 1 m resolution. Each scene contains 72 transmitter configurations (derived from 24 sampling rounds of 3 parameter sets) with \(|\mathcal{B}|=8\) candidate beams per configuration. For each configuration, a carrier frequency is randomly drawn from five bands (700 MHz, 1.8 GHz, 2.565 GHz, 3.5 GHz, and 4.9 GHz). All ground-truth RSRP maps are generated using Sionna RT, which results in a total of 72,000 configurations and 576,000 single-beam images. We adopt a split of 800/100/100 scenes for training/validation/testing, providing 57,600 training configurations. To ensure frequency balance, we construct Test500 by selecting exactly one configuration per carrier frequency from each test scene.
\begin{table}[H]
    \caption{Overall Prediction Performance on Test500 Subset}
    \vspace{-6pt}
    \label{tab:overall_comparison}
    \centering
    \footnotesize
    \setlength{\tabcolsep}{3pt}
    \begin{tabular}{@{}llrr@{}}
        \toprule
        Method & Prediction mode & MAE (dB) \(\downarrow\) & RMSE (dB) \(\downarrow\) \\
        \midrule
        RadioUNet~\cite{Levie2021RadioUNet} & Separate & 7.86 & 14.22 \\
        RME-GAN~\cite{Zhang2023RMEGAN} & Separate & 9.16 & 15.84 \\
        UNetDCN~\cite{Jaensch2026Coverage} & Separate & 7.39 & 13.04 \\
        RadioMamba~\cite{Jia2026RadioMamba} & Separate & 7.97 & 14.14 \\
        RadioFlow-Large-6ch~\cite{Jia2025RadioFlow} & Separate & 8.88 & 14.22 \\
        \midrule
        PathFinder~\cite{Zhong2026PathFinder} & Joint & 7.60 & 13.79 \\
        CKMTransUNet~\cite{Wang2025CKMTransUNet} & Joint & 8.31 & 14.15 \\
        \midrule
        \textbf{Proposed} & Separate & \textbf{6.70} & \textbf{12.59} \\
        \bottomrule
    \end{tabular}
\end{table}

We compute the MAE and root mean square error (RMSE) over valid pixels by excluding building interiors, invalid values, and sentinel no-signal pixels, and clipping the reference RSRP to \([P_c^{\mathrm{tx}}-300, P_c^{\mathrm{tx}}-30]\)~dBm. The predictor uses a DiT backbone (patch stride \(p_{\mathrm{patch}}=16\), hidden dimension \(d=1024\), 12 blocks, 8 attention heads, multilayer perceptron (MLP) ratio 4) with time and conditions injected via adaptive layer normalization. A spatial head (\(d_{\mathrm{head}}=32\), two \(3\times3\) convolutions) outputs raw parameter maps, which are constrained by the physical masks in Section~\ref{sec:esf_representation} to form \(\boldsymbol{J}_{c,t}\). We train with AdamW (batch size 32, peak learning rate \(2\times10^{-4}\), cosine decay) and select the best validation checkpoint. The SRP selection is configured with \(K=1024\), \(r_{\mathrm{reg}}=2\)~m, \(\rho_{\max}=10\), and \(\nu\in[1.5, 4.0]\) initialized at 2. The tilt beamwidth bounds are set to \([\sigma_{\vartheta,\min}, \sigma_{\vartheta,\max}] = [0.01, \pi/2]\)~rad. For the physical constraint terms in the training loss, the weighting coefficients are set as \(\lambda_{\text{amp}}=0.01\), \(\lambda_{\text{sp}}=1\), \(\lambda_{\text{wall}}=0.1\), \(\lambda_{\text{az}}=0.1\), \(\lambda_{\text{el}}=0.1\), and \(\lambda_{\text{ne}}=0.05\).

\begin{figure}[H]
    \centering
    \includegraphics[width=\columnwidth]{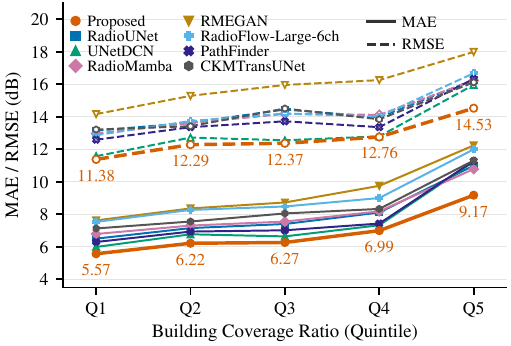}
    \vspace{-10pt}
    \caption{MAE/RMSE across building-coverage quintiles Q1--Q5 on Test500.}
    \label{fig:building_occupancy_comparison}
\end{figure}

\subsection{Comparison with Existing Methods}
\label{sec:comparison_existing_methods}
We compare our method against seven baselines, all evaluated on the full set of eight beam-specific maps. These baselines fall into two prediction paradigms. Five methods (RadioUNet~\cite{Levie2021RadioUNet}, RME-GAN~\cite{Zhang2023RMEGAN}, UNetDCN~\cite{Jaensch2026Coverage}, RadioMamba~\cite{Jia2026RadioMamba}, and RadioFlow-Large-6ch~\cite{Jia2025RadioFlow}) adopt a per-beam scheme, applying the same model independently to each beam with six-channel inputs and assembling the outputs. Our proposed method also follows this per-beam paradigm, predicting SRPs and rendering the RSRP map for each beam independently. In contrast, PathFinder~\cite{Zhong2026PathFinder} and CKMTransUNet~\cite{Wang2025CKMTransUNet} predict jointly from 13-channel inputs containing all eight beam-specific gain maps. All evaluations strictly follow the Test500 protocol and the evaluation rules defined in Section~\ref{sec:experimental_setup}.

Table~\ref{tab:overall_comparison} shows our method achieves the lowest MAE (6.70~dB) and RMSE (12.59~dB) among all seven baselines. Fig.~\ref{fig:building_occupancy_comparison} evaluates performance across building-coverage quintiles, where Q1 and Q5 represent the sparsest and densest building regions, respectively. While all methods degrade in dense areas (Q5) compared to sparse areas (Q1), ours consistently maintains the lowest errors across all five groups. Notably, its relative advantage is most pronounced in Q5 (increasing from 5.57/11.38~dB in Q1 to 9.17/14.53~dB in Q5), indicating that the SRP formulation effectively captures complex multipath effects (e.g., reflection and diffraction) that dominate in heavily obstructed scenarios.

Fig.~\ref{fig:maximum_rsrp_comparison} visualizes the maximum-RSRP maps for three cases (Q2, Q4, Q5), where our method achieves the lowest overall MAE. In Q2, it reduces the MAE from 19.35~dB (UNetDCN) to 14.27~dB by better reproducing the strong street corridor coverage. Similarly, in Q5, it improves from 11.89~dB (RadioMamba) to 8.02~dB by preserving signal strength between building rows. Conversely, in Q4, UNetDCN exhibits a lower local error within 80~m of the transmitter despite a higher map-level MAE (19.44~dB vs. 17.60~dB), and some sharp local variations are over-smoothed in our prediction. While these cases highlight specific visual improvements, the overall quantitative evaluation is established on the full Test500 benchmark.

\section{Conclusion}
\label{sec:conclusion}

This paper proposes a data- and physics-driven framework that predicts radio maps by learning geometry-constrained SRPs. A flow-matching predictor estimates these low-dimensional SRPs, which a closed-form physical renderer then translates into the final RSRP maps. This formulation fundamentally decouples wave propagation from environmental feature extraction, eliminating the need for ground-truth SRP labels. Extensive evaluations across multiple cities and frequencies demonstrate that our framework consistently outperforms seven baselines, yielding the lowest errors across all building-coverage quintiles, particularly in dense urban environments.

\end{document}